\documentclass[3p,times,procedia]{elsarticle}
\usepackage{nupha_ecrc}
\volume{00}
\firstpage{1}
\journalname{Nuclear Physics A}
\runauth{}
\jid{nupha}

\jnltitlelogo{Nuclear Physics A}

\usepackage{amssymb}
\usepackage[figuresright]{rotating}
\usepackage{amsfonts, amssymb, amsmath, graphicx, comment, bm, slashed, caption, subcaption, dcolumn,color}

\newcommand{\beq}{\begin{equation}}
\newcommand{\eeq}{\end{equation}}
\newcommand{\calH}{ {\cal H} }

\newcommand{\rmi}{ {\rm i} }

\begin{document}

\begin{frontmatter}

\dochead{}

\title{Phenomenological QCD equations of state for neutron stars 
}

\author[1,2]{Toru Kojo}
\author[3]{Philip D. Powell}
\author[1]{Yifan Song}
\author[1,4]{Gordon Baym}

\address[1]{Department of Physics, University of Illinois at Urbana-Champaign, 1110 W. Green Street, Urbana, Illinois 61801, USA}
\address[2]{Key Laboratory of Quark and Lepton Physics (MOE) and Institute of Particle Physics, Central China Normal University, Wuhan 430079, China}
\address[3]{Lawrence Livermore National Laboratory, 7000 East Avenue, Livermore, California 94550, USA}
\address[4]{Theoretical Research Division, Nishina Center, RIKEN, Wako 351-0198, Japan}

\begin{abstract}
We delineate the properties of QCD matter at baryon density $n_B=1-10n_0$ ($n_0$: nuclear saturation density), through the construction of neutron star equations of state that satisfy the neutron star mass-radius constraints as well as physical conditions on the speed of sound. The QCD matter is described in the 3-window modeling: at $n_B \lesssim 2n_0$ purely nuclear matter; at $n_B \gtrsim 5n_0$ percolated quark matter; and at $2n_0 \lesssim n_B \lesssim 5n_0$  matter intermediate between these two which are constructed by interpolation. Using a schematic quark model with effective interactions inspired from hadron and nuclear physics, we analyze the strength of interactions necessary to describe observed neutron star properties. Our finding is that the interactions should remain as strong as in the QCD vacuum, indicating that gluons at $n_B =1-10\,n_0$ remain non-perturbative even after quark matter formation.
\end{abstract}


\end{frontmatter}

\section{Neutron star constraints on the QCD equation of state}
\label{sec:1}
\begin{figure}[b]
\vspace{-0.2cm}
\begin{minipage}{0.5\hsize}
\vspace{-0.2cm}
\hspace{1.0cm}
\includegraphics[width = 0.75\textwidth]{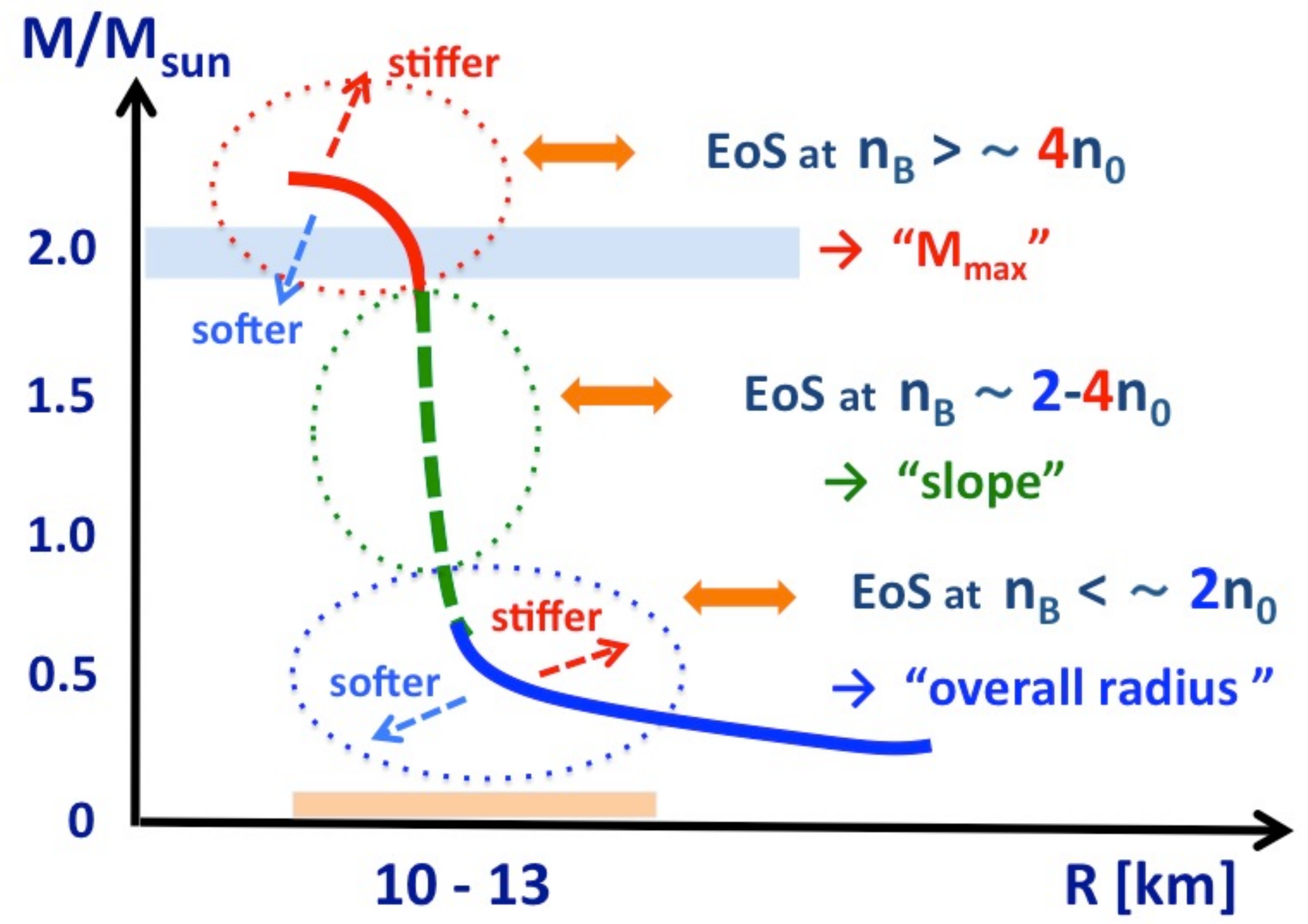}
\end{minipage}
\begin{minipage}{0.5\hsize}
\vspace{0.1cm}
\hspace{0.1cm}
\includegraphics[width = 0.72\textwidth]{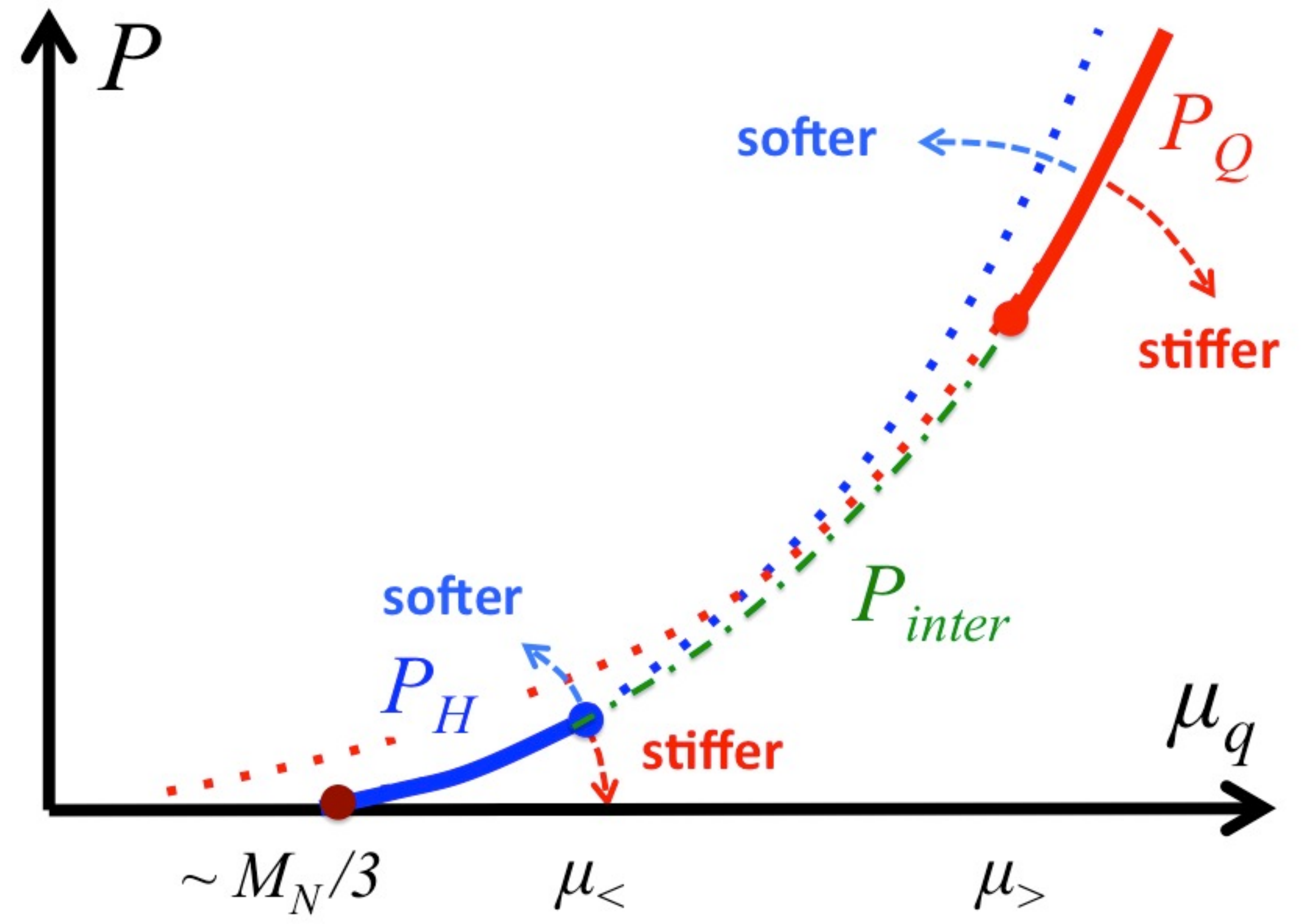}
\end{minipage}
\vspace{-0.2cm}
\caption{
\footnotesize{
(left) The correlation between the shape of the $M$-$R$ curve and pressures at several fiducial densities. (right) $P(\mu_q)$ curves. The 3-window modeling assumes that only the bold lines is trustable (the dotted lines are their extrapolations). The green curve is the interpolated pressure.
}
}
\vspace{-0.8cm}
\label{MR}
\vspace{-1.2cm}
\end{figure}

While RHIC and LHC have been excellent laboratories to study hot and high energy QCD, neutron stars are unique cosmic laboratories to study cold dense QCD. Solving  the Tolman-Oppenheimer-Volkoff equation for a given QCD equation of state, one can predict neutron star mass-radius ($M$-$R$) relations which are observable. This procedure is invertible \cite{Lindblom1992}; one can directly reconstruct the QCD equation of state once the $M$-$R$ relation is established from observation.  Although the $M$-$R$ curve is not determined precisely because of uncertainties in the radius determinations, current observations already provide tight constraints on equations of state. Our objective in this talk is to delineate properties of QCD matter, in light of equations of state inferred from current neutron star constraints.

As discussed by Lattimer and Prakash \cite{Lattimer:2006xb}, the features of the $M$-$R$ curve are determined by the pressures at different densities, see Fig.~\ref{MR} (left). The overall radii are correlated with equations of state at\footnote{
We use the notation: $n_B$, baryon density; $n_0 \simeq 0.16\,{\rm fm}^{-3}$, nuclear saturation density; $\mu_q$, quark chemical potential; $P$, pressure; $\varepsilon$, energy density; $c_s=\sqrt{\partial P/\partial \varepsilon}$, speed of sound); $T$, temperature; and $M_N$, nucleon mass. We take natural units, $c=\hbar=1$.} 
$n_B \lesssim 2n_0$; stiffer (softer) equations of state lead to larger (smaller) radii\footnote{
Stiffer equations of state have larger $P$ at given $\varepsilon$, not to be confused with $P$ at given $\mu_q$. See Fig.~\ref{MR} (right).}. At higher density the $M$-$R$ curve increases upward in the vertical direction without significant change in the radius; its slope is determined by the pressure at $2 n_0 \lesssim n_B \lesssim 4n_0$. The maximum mass is determined by the pressure at $n_B \gtrsim 4n_0$. In Fig.\ref{MR} (right) we also indicate the stiffness in terms of $P(\mu_q)$.

The above discussion can be combined with three constraints: (i) the existence of two neutron stars of two solar masses ($2M_\odot$)  \cite{2m_1} requires stiff equations of state at $n_B \gtrsim 4n_0$; (ii) the recent indications in neutron star radii analyses of rather small star radii $10-13\, {\rm km}$ \cite{Ozel2010,Steiner:2012xt,Ozel2015}, suggesting soft equations of state at $n_B \lesssim 2n_0$;  and (iii) thermodynamic and causality constraints on the speed of sound, $0 \lesssim c_s^2  \lesssim 1$.  While the analyses are not as precise as the mass determinations, soft equations of state at low density are actually consistent with Danielewitz's constraint \cite{Danielewicz:2002pu} obtained from heavy ion data and recent Monte-Carlo many-body calculations \cite{Gandolfi:2015jma}.

It is difficult to reconcile all these constraints simultaneously by constructing an equation of state that is soft at low density and stiff at high density. By definition, stiff equations of state have larger $P$ at given $\varepsilon$.  Thus the curve connecting these soft and stiff domains tends to contain rather large $\partial P/\partial \varepsilon$ and has the danger of violating constraint (iii). In this way, the three conditions impose constraints one another. If we ignored, for instance, the radius constraint (ii), then we have only to construct an equation of state which is stiff from low to high densities and thereby satisfy the constraints (i) and (iii). But in this work all these constraints are included in constructing equations of state.

\section{Three window modeling of QCD matter}
\label{sec:2}

\begin{figure*}
\hspace{2.50cm}
\includegraphics[width = 0.7\textwidth]{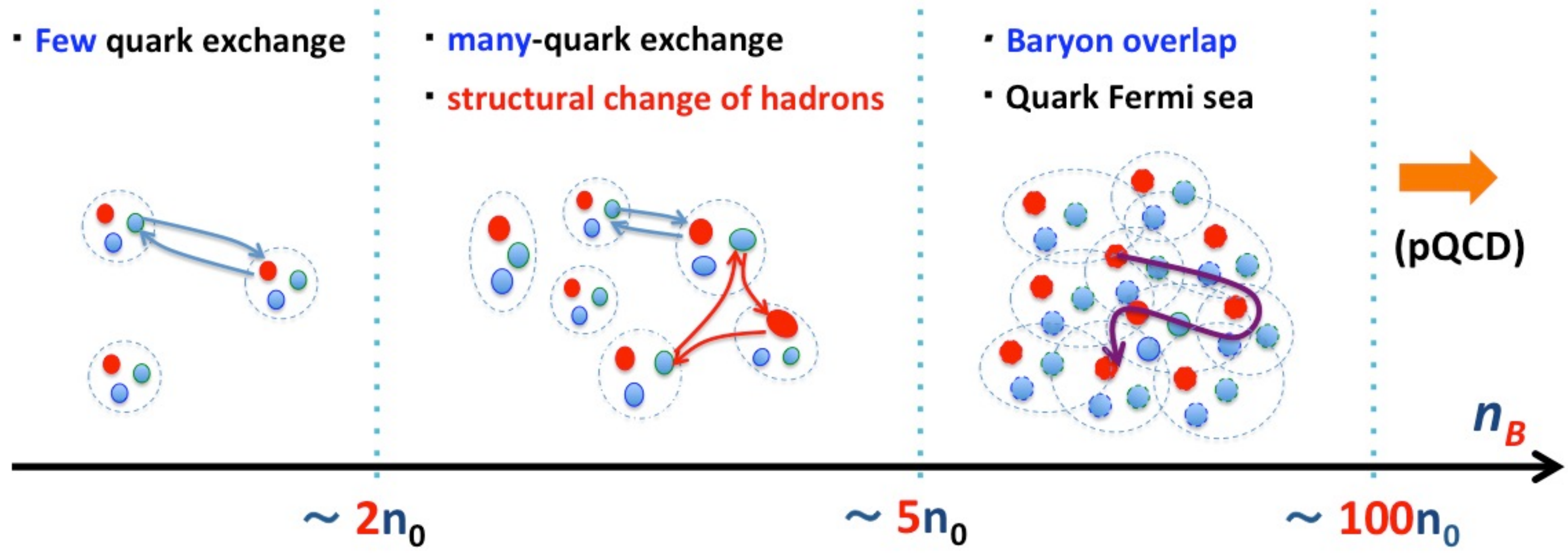}
\vspace{-0.3cm}
\caption{
\footnotesize{The 3-window description of QCD matter, see text.}
\vspace{-0.4cm}
}\label{fig:3-window}
\end{figure*}

Following Masuda-Hatsuda-Takatsuka \cite{Masuda:2012kf}, we consider the 3-window modeling of the QCD matter in which three domains are separately discussed (Fig.\ref{fig:3-window}): (a) At $n_B \lesssim \,2n_0$, nuclear matter is dilute and nucleons exchange only few mesons, allowing one to use sophisticated nuclear many-body calculations; (b) at $n_B \gtrsim \,2n_0$, nucleons start to exchange many mesons (or quarks) and many-body forces become increasingly important. With many quark exchanges it is natural to expect modifications of hadronic wavefunctions. It is also in this density region where hyperons begin to emerge, causing softening problems; (c) At $n_B \gtrsim \,5n_0$ baryon wavefunctions spatially overlap and quarks begin to travel around (percolation), forming quark matter. Nevertheless the matter is strongly correlated; in fact perturbative QCD (pQCD) calculations predict that the weak coupling expansion \cite{Kurkela:2009gj} does not converge well below $n_B \sim 100 \,n_0$, indicating that non-perturbative gluons are still important at densities relevant for neutron stars, $n_B = 1-10\,n_0$.

In practice, we use for the purely nuclear descriptions below $2n_0$, the SLy equation state for $n_B = 0-0.5n_0$ \cite{Douchin:2001sv} and the Akmal-Pandharipande-Ravenhall (APR) equation of state \cite{Akmal:1998cf} for $n_B = 0.5-2.0n_0$. For percolated quark matter above $5n_0$, we use quark models inspired from hadron and nuclear physics. In the intermediate region, we construct an equation of state by simply interpolating the APR and quark model equations of state. The interpolation variable is $P(\mu_q)$, and we match it to APR and quark pressures at $\mu_>$ and $\mu_<$ such that $n_B(\mu_<) = 2n_0$ and $n_B(\mu_>) = 5n_0$, respectively (e.g., Fig.\ref{MR} (right)). The matching conditions are imposed up to the second derivative of $P$ with respect to $\mu_q$. For details, see Refs. \cite{Kojo:2014rca}.

It should be emphasized that the 3-window modeling is quite different from the conventional hybrid modeling based on a Maxwell construction. In the latter, hadronic and percolated quark matter equations of state are considered as distinct, and as $\mu_q$ increases, the quark matter pressure $P_q(\mu_q)$ must intersect with the hadronic pressure $P_h(\mu_q)$ from below and then stay larger than $P_h(\mu_q)$. (This is necessary to have hadronic matter at low density and quark matter at high density.) For this to happen, the quark pressure must grow sufficiently fast as $\mu_q$ increases, but such equations of state are soft. Thus conventional modeling excludes stiff quark pressure curves. In particular, if we choose a soft hadronic equation of state, as implied by the radius constraint, it is very difficult to make quark equations of state stiff enough to satisfy the $2M_\odot$ constraint.

The conventional hybrid construction, however, has fundamental weakness; as we noted, purely hadronic and percolated quark matter pictures are valid in different domains so that it is not quite possible to reliably compare the two pressures. Especially it is not reasonable to use the {\it extrapolated} hadronic pressure at high density to reject stiff quark matter equations of state. In contrast, the 3-window construction does not utilize an extrapolated hadronic pressure, and one can construct stiff quark equations of state independent of the extrapolated hadronic pressure.  An example is shown in Fig.\ref{MR} (right). This allows us to consider quark equations of state that have not been fully explored.

It is instructive to quote an example where the difference between the conventional and 3-window constructions appear clearly: QCD at finite $T$. At low $T$ the QCD pressure is described well by a non-interacting hadron resonance gas (HRG) model. But at $T > T_c$, the pseudo-critical chiral phase transition temperature, the HRG pressure starts to overshoot the QCD pressure calculated in the lattice QCD. This occurs because near $T\sim T_c$ thermally excited hadrons begin to overlap and the interactions among them become important. On the other hand, at high $T$ pQCD calculations seem to work, but when we extrapolate the results toward low temperature, around $2-3T_c$ pQCD the pressure starts to overshoot the lattice QCD pressure; the lack of confining effects allows colored excitations to contribute to the pressure.

If we treated the HRG and pQCD pressures as real at all $T$ we would unrealistically find the deconfined phase at low $T$ and the hadronic phase at high $T$. This example illustrates the danger of trusting that  the extrapolated pressure describes real physics. On the other hand, the 3-window construction restricts the use of the HRG pressure to $T\lesssim T_c$ and pQCD pressure to $T \gtrsim 2-3T_c$, and then interpolates between them to reproduce the qualitatively correct result.   We expect a similar utility for cold dense QCD.

\section{Delineating QCD matter in a schematic quark model}
\label{sec:3}

\begin{figure}[b]
\vspace{-0.2cm}
\begin{minipage}{0.5\hsize}
\vspace{-0.cm}
\hspace{1.0cm}
\includegraphics[width = 0.65\textwidth]{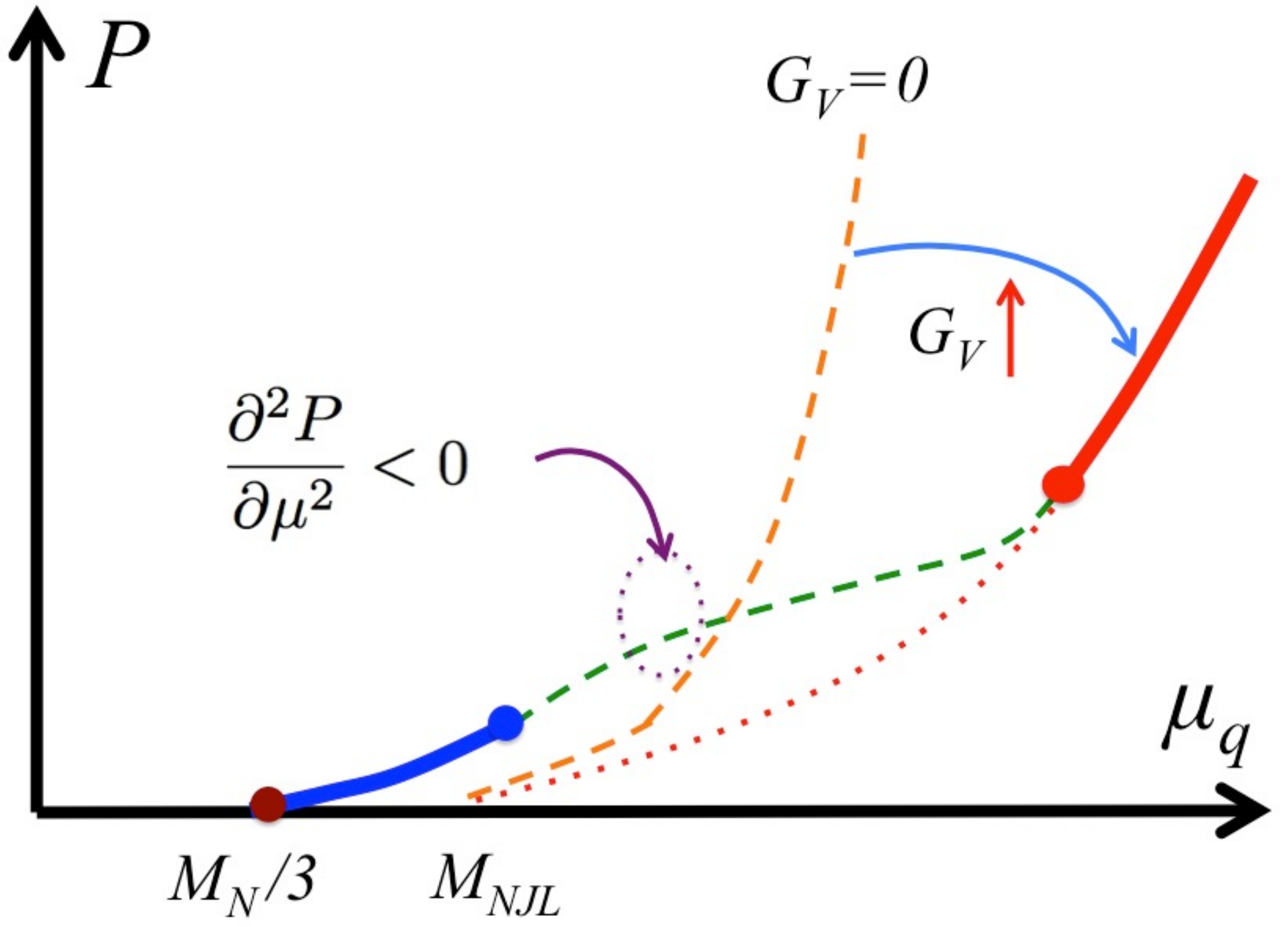}
\end{minipage}
\begin{minipage}{0.5\hsize}
\vspace{0.cm}
\hspace{0.cm}
\includegraphics[width = 0.65\textwidth]{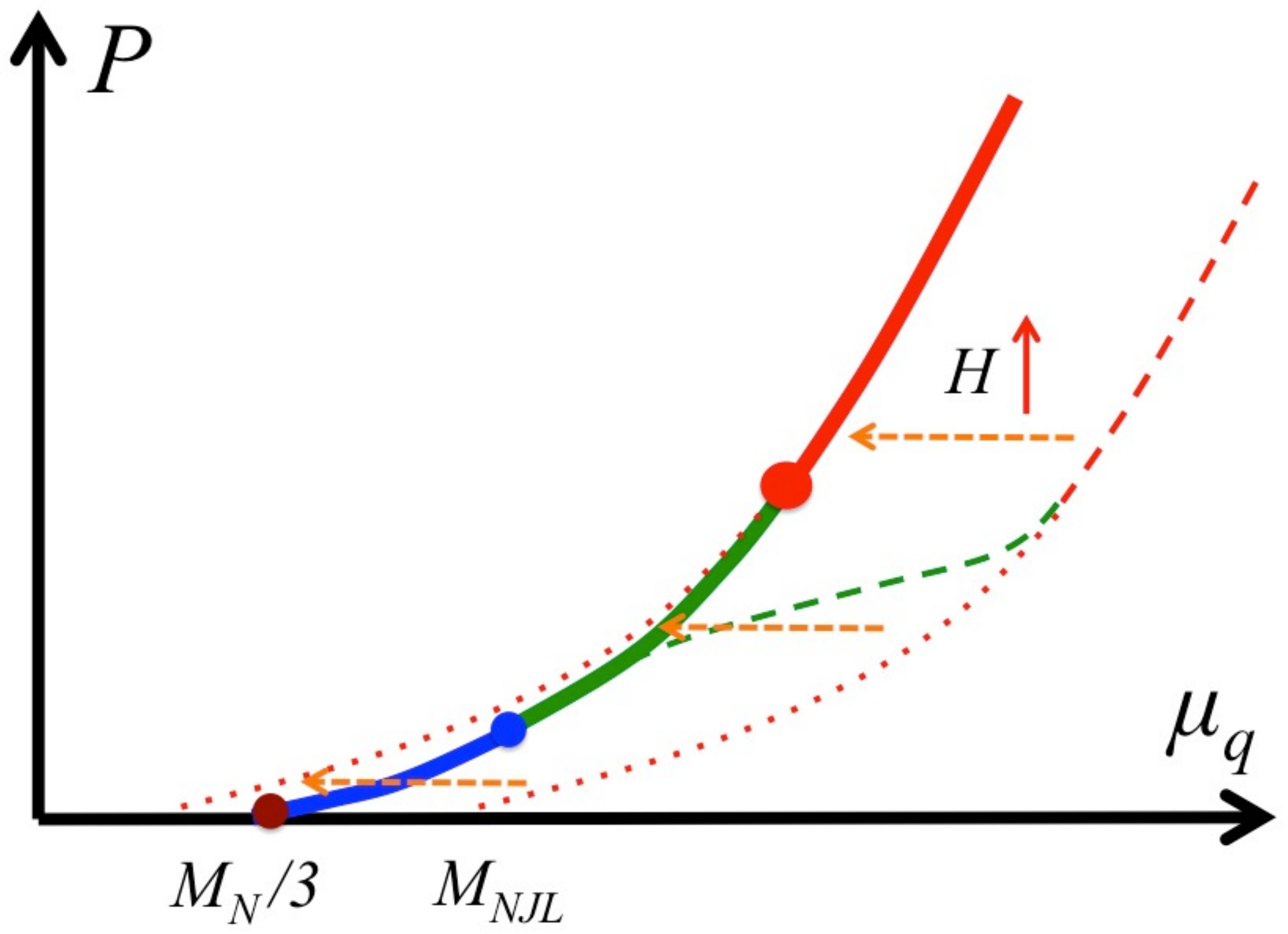}
\end{minipage}
\vspace{-0.2cm}
\caption{
\footnotesize{
(left) The stiffening by nonzero $G_V$, and the (unphysical) inflection point in the interpolated pressure. (right) Large $H$ shifts the pressure curve toward low $\mu_q$ region, erasing the inflection point. The resulting equation of state is physical and stiff.
}
}
\vspace{-0.8cm}
\label{NJL}
\vspace{-1.2cm}
\end{figure}

Now we consider quark models for the percolated region, referring to the ideas developed in the last sections. Our purpose here is to phrase the (supposed) neutron star equations in terms of the language developed for hadron and nuclear physics, which we expect to be suitable for the description of strongly correlated matter. We write the effective Hamiltonian for 3-flavors schematically as
\begin{align}
\calH  
& 
= \overline{q} (\rmi \gamma_0 \vec{\gamma}\cdot \vec{\partial} + m -\mu_q \gamma_0)q 
- \frac{G_s}{2} \sum^8_{i=0} \left[ (\overline{q} \tau_i q)^2 + (\overline{q} \rmi \gamma_5 \tau_i q)^2 \right] 
+ 8 K ( \det\,\!\!_{\rm f} \bar{q}_R q_L + \mbox{h.c.}) \nonumber \\
&~~~ + \calH_{ {\rm conf} }^{ {\rm 3q\rightarrow B} } 
 - H \!\sum_{A,A^\prime = 2,5,7} \!
 \left(\bar{q} \rmi \gamma_5 \tau_A \lambda_{A^\prime} C \bar{q}^T \right) \left(q^T C \rmi \gamma_5 \tau_A \lambda_{A^\prime} q \right) + \frac{G_V}{2} (\overline{q} \gamma^\mu q)^2
 \, ,\nonumber 
\end{align}
where $\tau_A$  and $\lambda_A$ are Gell-Mann matrices for flavors and colors respectively. The first line in the equation is the standard NJL model which describes the chiral symmetry breaking and restoration.  $\calH_{ {\rm conf} }^{ {\rm 3q\rightarrow B} }$ should describe the confining effects which trap 3-quarks into single baryon. But by restricting the use of our models within the percolated domain, we shall assume this term to be negligible. The term with coupling $H$ is the diquark-diquark or color-magnetic interaction which is responsible for e.g. the $N-\Delta$ splitting in hadron spectroscopy.  The last term is the repulsive vector interaction inspired by nuclear interactions mediated by the $\omega$-meson exchange. Adding leptons, this model is treated within the mean-field approximation, together with constraints of charge neutrality, $\beta$-equilibrium, and color neutrality. For details, see Refs.~\cite{Kojo:2014rca}.

We use the Hatsuda-Kunihiro parameter set for ($G_s, K$), the UV cutoff $\Lambda_{ {\rm NJL} }$, and current quark masses \cite{Hatsuda:1994pi}. The diquark and vector couplings ($H$, $G_V$) are treated as free parameters. Because of medium effects, in principle the parameters in the percolated domain can be different from their vacuum counterparts.  Our goal is to constrain these parameters at $n_B \gtrsim 5n_0$ by neutron star constraints, and then delineate properties of strongly correlated matter from the behaviors of these parameters \cite{Fukushima:2015bda}.

The impact of each term is as follows: (i) With the standard NJL model alone, our 3-window equations of state cannot pass the $2M_\odot$ constraint. (ii) Increasing $G_V$ from zero can increase the stiffness (Fig.~\ref{NJL} left). However, the large $G_V$ also introduces a problem when we interpolate between the low and high density pressures; the quark model pressure at larger $G_V$ tends to appear at higher $\mu_q$, so that the interpolating curve tends to have inflection points, and thus a region of negative curvature, $\partial^2 P/\partial \mu_q^2<0$. In this region the compressibility is negative, $c_s^2<0$, and the system is unstable; the interpolating curve must be regarded as unphysical. (iii) The problem of the negative curvature can be cured by increasing $H$ which shifts the quark pressure curve toward low $\mu_q$ (Fig.~\ref{NJL} right). This shift can be expected if one recalls the $N$-$\Delta$ splitting in the constituent quark models, where the attractive force reduces the constituent quark mass of $\gtrsim 330\, {\rm MeV}$ to (or below) one-third of nucleon mass.  Sufficiently large $H$ erases the inflection points. 

At $G_V \sim H \sim G_s$, one can construct  pressure curves that satisfy three conditions discussed in Sec.1. The radius is about $\sim 11.3\, {\rm km}$ which is mainly determined by the APR equation of state. The $2M_\odot$ constraint can be satisfied by taking $G_V \gtrsim 0.5G_s$. The thermodynamic and causality constraints require larger $H$ for larger $G_V$ to remove the inflection points;  $H$ should be $\sim 1.5G_s$ \cite{Kojo:2014rca}.

Note that the neutron star constraints require model parameters at $n_B\gtrsim 5n_0$ as large as the vacuum NJL coupling $G_s$. This is the main finding from our exercise.  We claim that this is a signal that the gluons remain non-perturbative as in the QCD vacuum, since quark models reflect the gluon dynamics in terms of effective couplings; if gluons become perturbative, the quark model couplings should be much smaller than the vacuum values, causing severe softening problems. These problems include chiral restoration at low $\mu_q$ and the necessity to add the gluon bag constant, both of which substantially soften equations of state. Thus we infer the existence of quark matter with non-perturbative gluons, which was conceptualized by McLerran and Pisarski \cite{McLerran:2007qj}.

In the present study we used smooth curves for the interpolation, assuming the crossover from hadronic to quark matter. This possibility is feasible especially when we demand neutron star radii to be small, $R\lesssim 13\,{\rm km}$, or equations of state to be soft at low density and stiff at high density. This condition disfavors strong bending in the interpolated pressure which would appear at a first order phase transition point. If the radii turn out to be large, $R\gtrsim 13\, {\rm km}$, one can allow very stiff hadronic pressure with which the quark pressure can remain stiff even after the first order phase transition \cite{Benic:2014jia}. The precise radius estimates, which still need confirmation, have direct implications on the nature of QCD phase diagram.

\section*{Acknowledgments}
T.K. thanks Kenji Fukushima for his collaborative work \cite{Fukushima:2015bda} which sharpened ideas about the medium couplings in effective models.
This work was supported in part by NSF Grants PHY09-69790 and PHY13-05891.

\bibliographystyle{elsarticle-num}

\end{document}